\pdfoutput=1
\documentclass[10pt,prl,aps,superscriptaddress,twocolumn]{revtex4-1}
\usepackage{amsmath,bbm}
\usepackage{amsthm}
\usepackage{amsmath}
\usepackage{latexsym}
\usepackage{amssymb}
\usepackage{float}
\usepackage{graphicx}   
\usepackage{color}
\usepackage{mathpazo}
\usepackage{comment}
\usepackage{enumitem}
\usepackage{multirow}
\usepackage{subcaption}
\usepackage{ragged2e}
\DeclareCaptionJustification{justified}{\justifying}
\usepackage[colorlinks=true,linkcolor=blue,citecolor=magenta,urlcolor=blue]{hyperref}
\newcommand{\be}{\begin{equation}}
\newcommand{\ee}{\end{equation}}
\newcommand{\bea}{\begin{eqnarray}}
\newcommand{\eea}{\end{eqnarray}}

\newcommand{\ket}[1]{|#1\rangle}

\makeatletter
\newtheorem*{rep@theorem}{\rep@title}
\newcommand{\newreptheorem}[2]{%
\newenvironment{rep#1}[1]{%
 \def\rep@title{#2 \ref{##1}}%
 \begin{rep@theorem}}%
 {\end{rep@theorem}}}
\makeatother

\newtheorem{thm}{Theorem}
\newreptheorem{thm}{Theorem}

\newtheorem{coro}{Corollary}

\begin{document}



\title{Threshold ($Q, P$) Quantum Distillation}


\author{Shashank Gupta}
\email{shashankg687@gmail.com}
\affiliation{Okinawa Institute of Science and Technology Graduate University, Okinawa, Japan}

\author{William John Munro}
\affiliation{Okinawa Institute of Science and Technology Graduate University, Okinawa, Japan}

\author{Carlos Cid}
\affiliation{Okinawa Institute of Science and Technology Graduate University, Okinawa, Japan}
\affiliation{Simula UiB, Bergen, Norway}
%



\begin{abstract}
Quantum distillation is the task of concentrating quantum correlations present in $N$ imperfect copies 
using free operations by involving all $P$ parties sharing the quantum correlations. We present a threshold quantum distillation task where the same objective is achieved but using fewer parties $Q$. In particular, we give exact local filtering operations by the participating parties sharing a high-dimension multipartite GHZ or W state to distil the perfect quantum correlation. Specifically, an arbitrary GHZ state can be distilled using just one party in the network, as both the success probability of the distillation protocol and the fidelity after the distillation are independent of the number of parties. However, for a general W-state, at least $(P-1)$ parties are required for the distillation, indicating a strong relationship between the distillation and the separability of such states. Further, we connect threshold entanglement distillation and quantum steering distillation.
\end{abstract}


\maketitle


\textit{Introduction---} The idea of a threshold is fundamental in classical cryptography to ensure privacy, security, and fault tolerance \cite{Desmedt2011}. A well-celebrated protocol underpinning the threshold concept is the Shamir threshold ($Q, P$) secret-sharing scheme, where any $Q$ out of $P$ parties can collaborate to unravel a hidden secret \cite{Shamir1979}. Several protocols have adopted threshold features to improve secure multi-party computation, signatures, and database management in classical cryptography \cite{DeSantis1994, Bogdanov2012, Cramer2015, Evans2018, Archer2018, Kamm2024}. In quantum cryptography, the threshold concept is crucial as the viability hinges on the quantum resource exceeding a specific threshold bound \cite{Briegel98, Campbell13, Tan20, Rengaswamy2024}. To achieve this, the distillation of the quantum resources is fundamental \cite{Sun22, Yan22, Wei2022, MiguelRamiro2023}, leading to the proposal of numerous distillation or quantum privacy amplification protocols. In the bipartite systems, distillation protocols for quantum entanglement \cite{Bennet96,Ekert96,Horo98,Horo99,Horod99,Horo01}, EPR steering \cite{Nery20,Liu20, Hao2024} and Bell nonlocality \cite{Forster09,Forster11,Peter10,Brunner11,Wu2013,Hyer2013} are well established.  Recent research have addressed a long-standing issue in two-qubit systems by closing the Werner gap for quantum steering \cite{Zhang2024}. Subsequent work has underscored the practical challenges and future directions of EPR steering \cite{Xiang2022}, underscoring its potential in quantum technologies \cite{Lee2023, Li2024}. In the context of multipartite systems, distillation protocols have been proposed for genuine entanglement \cite{Huang14, Ben08}, genuine steering \cite{Gupta21} and Bell nonlocality \cite{Wu10, Ye12, Ebbe2013, Pan2015}. However, most of these proposed distillation protocols are not practically efficient as they require the participation of all parties involved, either directly or indirectly, to extract ideal quantum resources. Additionally, the relationship between different distillation protocols is not clearly defined.
We propose a more efficient distillation protocol that requires the participation of fewer parties to achieve the desired outcome. Furthermore, we establish a robust correlation between the distillation of steering and entanglement. This proposal aims to enhance the practicality and understanding of quantum resource extraction in multipartite systems.


 \textit{Perfect quantum correlation---} That represents an extremal point in a given polytope, which is the most desirable in any quantum information processing task. However, environmental interactions and experimental imperfection obstruct one from realizing the perfect quantum correlation, thus degrading the performance of the implemented task. Distillation addresses this problem by concentrating on imperfect correlations. This letter focuses on distilling the underlying state or assemblage rather than specific correlation. Once we distil the state, we distil the correlation as well.

\begin{figure}[h!]
	\includegraphics[width=0.45\textwidth, height=5cm]{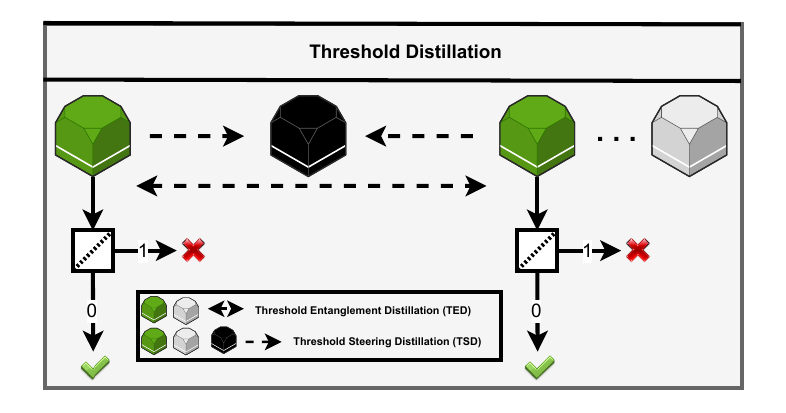}
\caption{ \footnotesize{Threshold Distillation (TD) framework. Green-coloured parties participate (PP) in the distillation protocol by applying local filtering operations (shown as filters). If the subsystem undergoes operation-0, the distillation protocol is successful; otherwise, it fails for the respective run. The non-participating parties (NP) act as ideal parties and are marked grey in colour. There may be some uncharacterized parties (black coloured) in the network as well; such an asymmetric network comes under the framework of quantum steering. Classical communication is two-way (double-sided dashed arrow) between the characterized parties and one-way (single-sided dashed arrow) from the characterized party to the uncharacterized party. PP, NP, and two-way communication constitute TED, whereas PP, NP, Uncharacterized parties and one-way communication correspond to TSD.}}
	\label{TD}
\end{figure}

\textit{Threshold quantum distillation ($N, d, P, Q$)---}
Consider a quantum network constituted by $P$ spatially separated parties. Let the parties share $N$ copies of quantum states (or assemblage in case of quantum steering) $\rho \in \mathcal{B}(\otimes_{i=1}^P \mathcal{H}_i^d)$ where $d$ is the dimension of each subsystem. $\mathcal{B}$ is the set of all density matrices in Hilbert space $\otimes_{i=1}^P \mathcal{H}_i^d$ (see Fig. (\ref{TD})). In general, the dimension of each subsystem might be different, but we consider it to be the same for simplicity. The objective of the threshold quantum distillation is to extract lesser copies of the perfect quantum states (or assemblage) using a smaller number $Q$ $(Q<P)$ of parties and free operations only. The free operations are defined with respect to the distillable resource. For example, local operation and classical communications (LOCCs) are used for entanglement, and one-way LOCCs are used for quantum steering.

%
\textit{Performance Metrics---}
The performance of the threshold distillation is measured using a standard fidelity metric. The fidelity measure is state (assemblage) fidelity for entanglement (steering) respectively with respect to the perfect state or assemblage. Let $\rho^{\text{dist}}$ be the state after the application of the threshold distillation protocol, and $\rho^{\text{perf}}$ is the perfect state in the same Hilbert space. The state fidelity between them
is {\small{$\mathcal{F}_s (\rho^{\text{dist}}, \rho^{\text{perf}}):= \text{max}_{\text{perf}} \Bigg[\text{Tr}\Big(\sqrt{\sqrt{\rho^{\text{dist}}} \rho^{\text{perf}} \sqrt{\rho^{\text{dist}}}}\Big)\Bigg]^2$}}. Similarly, Let $\{\sigma_{a_i|x_i}^{\text{dist}}\}$ be the assemblage after the application of the threshold steering distillation protocol and $\{\sigma_{a_i|x_i}^{\text{perf}}\}$ is the assemblage of the same dimension with perfect steerability. The assemblage fidelity between them
is  $\tiny{\mathcal{F}_a (\{\sigma_{a_i|x_i}^{\text{dist}}\}, \{\sigma_{a_i|x_i}^{\text{perf}}\}):= \text{min}_{x_i} \sum_{a_i} \Bigg[\text{Tr}\Big(\sqrt{\sqrt{\sigma_{a_i|x_i}^{\text{dist}}} \sigma_{a_i|x_i}^{\text{perf}}\sqrt{\sigma_{a_i|x_i}^{\text{dist}}}}\Big)\Bigg]^2}$. Note that minimization is required for the measurement settings of the uncharacterized parties \cite{Nery20}.

\textit{Threshold entanglement distillation (TED) protocol ($N, d, P, Q$)---} In a quantum network consisting of $P$ parties sharing imperfect entanglement, the idea of TED is that some lesser number of parties ($Q$) are mentioned as participating parties (PP) perform local filtering operation to distil the perfect entanglement with some success probability and non-participating parties (NP) do nothing (see Fig. {\ref{TD}}). We start with $N \geq 2$ copies of initial partially entangled states $\ket{\psi^{\text{ini}}}$. Each of the PP perform dichotomic POVM $\mathcal{G}^i := \{G_0^i, G_1^i\}$, satisfying $G_{o_i^u}^i \geq 0 \forall o_i^u \in \{0,1\}$ and $G_0^i+G_1^i = 1$ on the $u^{th}$ copy of the initial state with ($u \in \{1, 2, \cdots, N-1\}$). If all the PP applied $o_i^u = 1 \forall u$, then the PP set $o_i^N = 0$ for the last copy, otherwise set $o_i^N$ = 1 for the last copy (this will ensure that even if the distillation protocol is failed, the network has at least one copy with the initial correlation). Next, the PP send the string $o_i:=\{o_i^1, o_i^2, o_i^3, \cdots, o_i^N\}$ to all the parties in the network through the classical channel.

Finally, all the parties discard the $u^{th}$ copies for which $o_i^u  = 1$. The remaining copies of the states give the protocol output after the PP's filtering operations.

Next, for the filtering operation, we introduce the $\quad$ following notation $K_{o_i^u}^i = \sqrt{G_{o_i^u}^i}$ such that $G_{o_i^u}^i = K_{o_i^u}^{i^{\dagger}}K_{o_i^u}^i$. In the above protocol, the state of the $u^{th}$ copy after the application of the local filtering operations by the PP is
\begin{equation}
    \rho_u^{\text{dist}} = \frac{\otimes_{PP} K_{o_i^u}^{PP} \otimes_{NP} \mathcal{I}^{NP}. \rho_u^{\text{init}}. \otimes_{PP} K_{o_i^u}^{{PP}^{\dagger}} \otimes_{NP} \mathcal{I}^{NP}}{\text{Tr}[\otimes_{PP} K_{o_i^u}^{PP} \otimes_{NP} \mathcal{I}^{NP}. \rho_u^{\text{ini}}. \otimes_{PP} K_{o_i^u}^{{PP}^{\dagger}} \otimes_{NP} \mathcal{I}^{NP}]} \nonumber
\end{equation}
The success probability of the distillation protocol for the $u^{th}$ copy ($P_s^u$) $ = \text{Tr}[\otimes_i K_{0}^P \otimes_j \mathcal{I}^{NP}. \rho^{\text{init}}. \otimes_i K_{0}^{P^{\dagger}} \otimes_j \mathcal{I}^{NP}]$. The action of the local filtering operation is independent on each copy, so the failure probability (any PP gets outcome $1$ for the $u^{th}$ copy where $u \in \{1, 2, \cdots, N-1\}$) is $(1-P_s^u)^{N-1}$. This implies that the overall success probability of the threshold distillation protocol ($P_s$) is $1-(1-P_s^u)^{N-1}$. Now, the final state ($\rho^{\text{dist}}$) after the distillation protocol can be written as a convex combination of the perfect state with probability $P_s$ and the initial state with probability $1-P_s$. This state is used to determine the state fidelity.

\textit{Threshold steering distillation (TSD) protocol (N, d, S, P, Q)---} In a network consisting of characterized and uncharacterized (S) parties sharing entanglement. The measurements by the uncharacterized parties can influence the measurement outcome of the characterized parties, giving rise to the phenomenon of quantum steering. The idea behind TSD is that some characterized parties act like the PP and apply the local filtering operation on their part of the assemblage (unnormalized conditional states left on the characterized parties subsystem after uncharacterized parties measurement). The operation performed by the characterized parties in the steering scenario is one-way LOCC (see Fig. (\ref{TD})). We start with $N \geq 2$ copies of initial partially steerable assemblages $\{\sigma^{\text{ini}}_{a_i|x_i}\}$. Each of the participating characterized parties perform dichotomic POVM $\mathcal{G}^i := \{G_0^i, G_1^i\}$, satisfying $G_{o_i^u}^i \geq 0 \forall o_i^u \in \{0,1\}$ and $G_0^i+G_1^i = 1$ on the $u^{th}$ copy of the assemblage with ($u \in \{1, 2, \cdots, N-1\}$). If all the PP applied $o_i^u = 1 \forall u$, then the PP set $o_i^N = 0$ for the last copy; otherwise, set $o_i^N$ = 1 for the last copy. Next, the PP sends the string $o_i:=\{o_i^1, o_i^2, o_i^3, \cdots, o_i^N\}$ to all the parties in the quantum network through the $\quad$ classical channel. Finally, all the parties discard the $u^{th}$ copy for which $o_i^u  = 1$. The remaining copies of the assemblages give the protocol output after the PP filtering operations.

In the above protocol, the state of the $u^{th}$ copy after the application of the local filtering operations by the participating characterized parties is
\begin{equation}
    \sigma^{\text{dist}}_{a_i|x_i} = \frac{\otimes_{PP} K_{o_i^u}^{PP} \otimes_{NP} \mathcal{I}^{NP}.\sigma^{\text{ini}}_{a_i|x_i}.\otimes_{PP} K_{o_i^u}^{{PP}^{\dagger}} \otimes_{NP} \mathcal{I}^{NP}}{\text{Tr}[\otimes_{PP} K_{o_i^u}^{PP} \otimes_{NP} \mathcal{I}^{NP}.\rho_{ch}^{\text{ini}}. \otimes_{PP} K_{o_i^u}^{{PP}^{\dagger}} \otimes_{NP} \mathcal{I}^{NP}]} \nonumber
\end{equation}
Here, $\rho_{ch}^{\text{ini}}$ is the conditional state on the characterized parties subsystems, i.e. $\rho_{ch}^{\text{ini}} = \text{Tr}_{unch}\rho^{ini}$. The success probability of the TSD protocol for the $u^{th}$ copy ($P_s^u$) is $\text{Tr}[\otimes_i K_{0}^P \otimes_j \mathcal{I}^{NP}. \rho_{char}^{\text{ini}}. \otimes_i K_{0}^{P^{\dagger}} \otimes_j \mathcal{I}^{NP}]$. The action of the local filtering operation is independent on each copy, so the failure probability (any PP gets outcome $1$ for the $u^{th}$ copy where $u \in \{1, 2, \cdots, N-1\}$) is $(1-P_s^u)^{N-1}$. This implies that the overall success probability of the TSD protocol ($P_s$) is $1-(1-P_s^u)^{N-1}$. Now, the final assemblage ($\{\sigma^{\text{dist}}_{a_i|x_i}\}$) after the distillation protocol can be written as a convex combination of the perfect assemblage with probability $P_s$ and the initial assemblage with probability $1-P_s$. This assemblage leads us to the assemblage fidelity.

\begin{thm}\label{thm_GHZ_TED}
(\textit{TED of the GHZ states}) In a quantum network sharing multipartite high-dimensional GHZ states of the form
\begin{equation}
  \ket{\psi^{ini}_{GHZ}} = \sum_{i=0}^{d-1} \alpha_i \ket{\otimes_{j=0}^{P-1} i_j}_d, 
  \label{GHZ_state}
\end{equation}
 quantum entanglement can be distilled using a TED protocol ($N, d, P, 1 \leq Q \leq P-1$) with the success probability per copy ($P_s^u = d \alpha_0^2$) and the state fidelity 
 {\small{
 \begin{equation}               \mathcal{F}_s(\rho^{\text{dist}},\rho_{GHZ}^{\alpha_i = \frac{1}{\sqrt{d}}}) = \Big[1 - \frac{1}{d} (1-P_s^u)^{N-1} (d- (\sum_{i=0}^{d}\alpha_i)^2)\Big],
    \label{GHZ_fidelity_eq}
 \end{equation}}}
  which are independent of the number of parties in the network and the number of PP.
\end{thm}

\begin{figure}[h!]
\begin{subfigure}[h]{0.49\linewidth}
	\includegraphics[width=\linewidth]{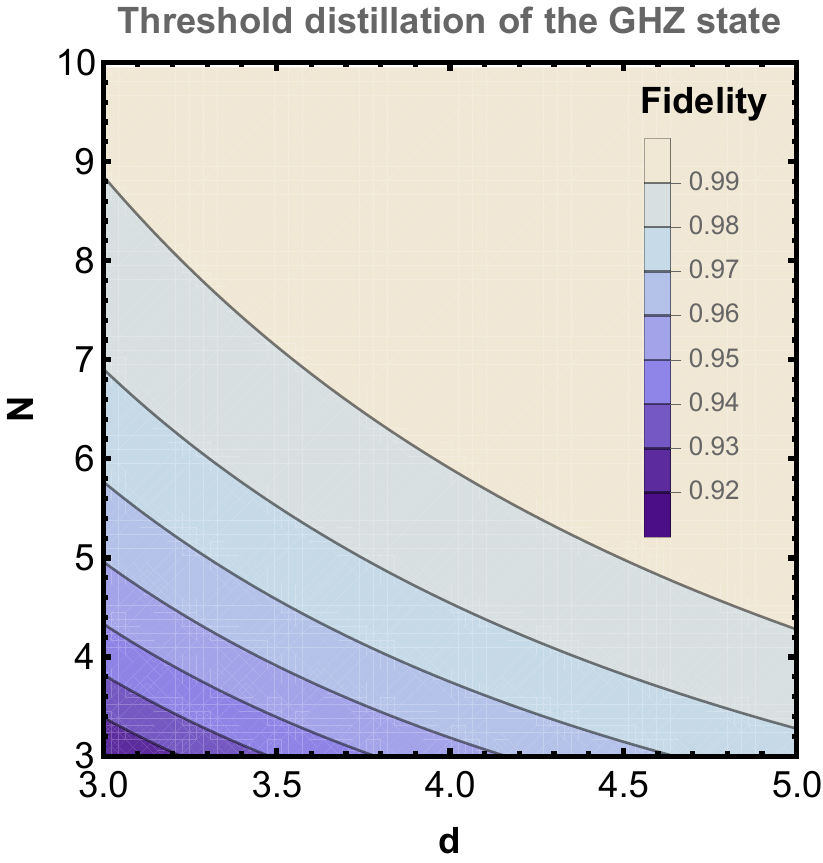}
\end{subfigure}
\begin{subfigure}[h]{0.49\linewidth}
    \includegraphics[width=\linewidth]{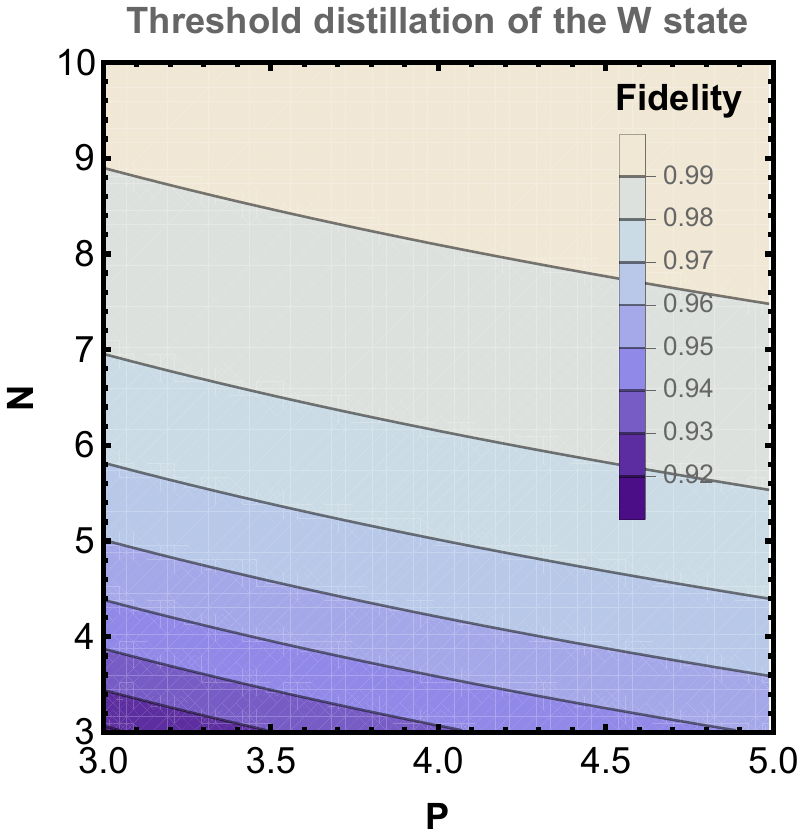}
\end{subfigure}
 \caption{\footnotesize{The performance of the threshold distillation of the GHZ state (\ref{GHZ_state}) on the left and W-state (\ref{W_state}) on the right is shown. The fidelity (\ref{GHZ_fidelity_eq}) is plotted concerning the number of copies $N$ and the dimension $d$ of the subsystem for a fixed value of the first state parameter ($\alpha_0 = 1/\sqrt{10}$) and the fixed difference between $d$ and the square of the sum of the state parameters (0.5). Similarly, the fidelity (\ref{W_fidelity_eq}) is plotted as a function of the number of copies $N$ and the number of the parties $P$ for a fixed value of the success probability ($P_s^u = 0.3$) and the fixed difference between $P$ and square of the sum of all the state parameters (0.5).}} 
 \label{contour}
\end{figure}

Using the local filtering operations by the PP ($1\leq Q\leq P-1$) on the initial state (\ref{GHZ_state}) and post-selecting the state when all the PPs get the outcome zero yields the expression of the fidelity (\ref{GHZ_fidelity_eq}). For the expression of Krauss operators for a specific $Q$, detailed proof, and toy examples, see supplementary materials (I) \cite{sm}. The fidelity improves remarkably with increasing $N$ and $d$ (see Fig. (\ref{contour})). The fidelity is more than 0.99 for more than $50\%$ of the region. It asymptotically approaches 1.  

Next, consider the assemblage derived from the state (\ref{GHZ_state}) by the action of the uncharacterized parties ($1 \leq S \leq P-1$) in two mutually unbiased bases (MUBs) \cite{Designolle2021}. Such assemblages are imperfectly steerable as the underlying state is imperfect. Only characterized parties can perform the local filtering operation to distil the assemblage; therefore, threshold distillation comes naturally in a realistic network. 

\begin{thm}\label{thm_GHZ_TSD}
(TSD of the GHZ assemblages) A quantum network sharing multipartite high-dimensional assemblages derived from the GHZ state (\ref{GHZ_state}) on the application of MUBs by uncharacterized parties,
quantum steering can be distilled using a TSD protocol ($N, d, 1 \leq S \leq P-2, P, 1 \leq Q \leq P-S-1$) with the assemblage fidelity same as the state fidelity \Big($\mathcal{F}_s(\rho^{\text{dist}},\rho_{GHZ}^{\alpha_i = \frac{1}{\sqrt{d}}})$\Big) (\ref{GHZ_fidelity_eq}).   
\end{thm}
In a given steering scenario ($1\leq S\leq P-2$), some of the characterized parties ($P-S$) can act as PPs ($1\leq Q \leq P-S-1$). The PPs use the same Krauss operators as in the case of TED to distil the assemblage. The PPs perform local filtering operations on their part of the assemblage and post-select the assemblage when all the PPs get the outcome-0. The expression of the assemblage fidelity is then derived by using distilled assemble and the perfect assemblage corresponding to the GHZ state ($\forall \alpha_i=1/\sqrt{d}$), where a minimization over the input measurements of the characterized parties is done. Note that $S=P-1$ is a valid steering scenario, and distillation of steering is possible using the only characterized party left. However, it is not a valid TSD as all the characterized parties are used for the distillation. See supplementary materials (II) \cite{sm} for detailed proof and toy examples.

\begin{figure}[h!]
	\includegraphics[width=0.24\textwidth, height=5cm]{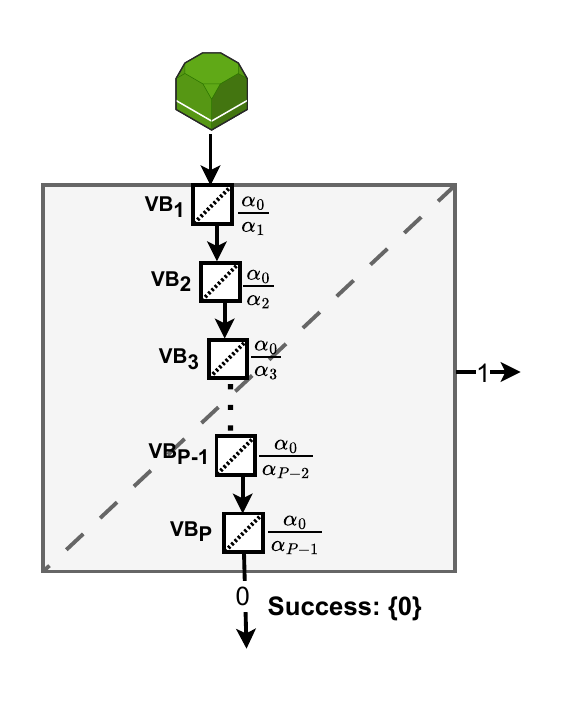}
	\caption{\footnotesize{The single party Krauss operator for the TED (TSD) of the GHZ state (\ref{GHZ_state}) (assemblage) can be realized using cascaded variable beamsplitters (VBs). The reflectivity coefficient is tuned as per the ratio of the state parameters concerning the first parameter ($\alpha_0$). The concept experiment follows from the recent experiment \cite{Nery20, Martnez2022}.}}
	\label{concept_exp}
\end{figure}

We know that the GHZ state (\ref{GHZ_state}) is a separable state as soon as one of the parties performs a projective measurement. This means the entanglement of GHZ type is tunable by just one party. Therefore, there exists, and we have found the Krauss operator of a single party to distil entanglement and steering of the GHZ kind. We propose a concept experiment to realize the Krauss operator of the single party to distil GHZ state (\ref{GHZ_state}) \cite{Nery20} (See Fig. (\ref{concept_exp})). The Krauss operator can be realized using a cascade of tunable beamsplitters (TBs) to tune the reflectivity coefficient of a TB as per the ratio of the state parameter concerning the first state parameter. Further tuning and post-selection are done using other passive optical components.  

\begin{coro}\label{rst_1_GHZ}
    A TED (TSD) is possible when the shared state (assemblage) in the quantum network is a multipartite high-dimensional non-maximally entangled GHZ state (assemblage), implying that GHZ state (assemblage) distillation and separability are strongly connected.
\end{coro}

We know that quantum steering is a stronger correlation than entanglement \cite{Jones07}. This means a steerable state is always entangled. This implies that if we devise a steering distillation protocol, it will also distil entanglement. 

\begin{coro}\label{rst_2_GHZ}
    The existence of the TSD in the quantum network sharing the assemblage derived from the GHZ state of the form (\ref{GHZ_state}) always indicate the existence of TED. The uncharacterized parties in the TSD can be non-participating parties in TED. 
\end{coro}
The above result holds as the number of non-participating parties in the entanglement scenario can be greater than or equal to the number of uncharacterized parties in the steering scenario for the GHZ-state (\ref{GHZ_state}), i.e. ($NP \geq S$).
Next, we consider the W-state under the action of threshold distillation.
\begin{thm}\label{thm_w_TED}
(TED of the W states) A quantum network sharing multipartite W states of the form
\begin{equation}
  \ket{\psi^{init}_{W}} = \sum_{i=0}^{P-1} \beta_i \ket{\cdots0_{i-1}1_i0_{i+1}\cdots}, 
  \label{W_state}
\end{equation}
 quantum entanglement can be distilled using a TED protocol ($N, d=2, P, Q = P-1$) with the success probability per copy \Bigg($P_s^u = P \frac{\Pi_{i=0}^{P-1}\beta_i^2}{\beta_{P-1}^{2(P-1)}}$\Bigg) and the state fidelity 
 {\small{
 \begin{equation}   \mathcal{F}_s(\rho^{\text{dist}},\rho_{W}^{\beta_i = \frac{1}{\sqrt{P}}}) = \Big[1 - \frac{1}{P} (1-P_s^u)^{N-1} (P- (\sum_{i=0}^{P-1}\beta_i)^2 )\Big],
     \label{W_fidelity_eq}
 \end{equation}}}
\end{thm}
Using the local filtering operations by the PP ($Q=P-1$) on the initial state (\ref{W_state}) and post-selecting
the state when all the PP get the outcome zero yields
the expression of fidelity (\ref{W_fidelity_eq}). For the exact expression of the
Krauss operators, detailed proof, and
toy examples, see supplementary materials
(III) \cite{sm}. The fidelity improves remarkably with increments in the number of copies ($N$) and the number of parties ($P$) (see Fig. (\ref{contour})), and it asymptotically approaches 1. 


Now, consider the assemblage derived from the W-state (\ref{W_state}) by the action of the first party in the two MUBs (computational and Hadamard basis). These are imperfect assemblages as the underlying state is imperfect. In this case, only characterized parties $P-1$ can perform the local filtering operations to distil the assemblage. Therefore, the threshold distillation naturally comes in a realistic network comprising characterized and uncharacterized parties. In this case, only steering distillation is possible and not threshold because at least $P-1$ parties are required to distil the assemblage derived from W-state (\ref{W_state}). For such a general state, distillation for other steering scenarios ($S>1$) is not feasible. However, distillation might be possible by putting constraints on the state parameters (equality of the state parameters).

\begin{thm}\label{thm_W_TSD}
(Steering Distillation (SD) of the W assemblages) For a quantum network sharing multipartite assemblages derived from the W state (\ref{W_state}) on the action of MUBs by uncharacterized parties, quantum steering can be distilled using an SD protocol ($N, d=2, S=1, P, Q=P-1$)  with the assemblage fidelity same as the state fidelity \Big($\mathcal{F}_s(\rho^{\text{dist}},\rho_{W}^{\beta_i = \frac{1}{\sqrt{P}}})$\Big) (\ref{W_fidelity_eq}).   
\end{thm}

In the one-sided device independent (1SDI) steering scenario ($S=1$), all the $P-1$ characterized parties use the same Krauss operators as in the case of TED to distil the assemblage. The PPs perform local filtering operations on their part of the assemblage and post-select the assemblage when all the PPs get the outcome $0$. The expression of the assemblage fidelity is then determined using distilled assemblage and the perfect assemblage ($\forall \beta_i = 1/\sqrt{P}$), where a minimization over the input measurements of the first party is done to determine the correct assemblage fidelity. For a detailed proof and toy example, see supplementary materials (IV) \cite{sm}.

The W state (\ref{W_state}) remains entangled up to the bipartite configuration. This means the genuine entanglement of W type is tunable by the collaboration of all the parties (less one due to the normalization). Therefore, there exists, and we have found the Krauss operator of $P-1$ parties to distil entanglement and steering of the W kind.

\begin{coro}\label{rst_1_W}
    A TED (SD) is possible with at least $P-1$ parties when the shared state (assemblage) in the quantum network is a multipartite non-maximally entangled W state (assemblage), implying that W state (assemblage) distillation and separability are strongly connected.
\end{coro}

It is understood that the correlation of quantum steering surpasses that of entanglement, as stated in \cite{Jones07}. Consequently, a state that exhibits steering is invariably entangled. Therefore, if a protocol for distilling steering is established, it will invariably result in the distillation of entanglement as well.

\begin{coro}\label{rst_2_W}
    The existence of the SD in the quantum network sharing the assemblage derived from the W state of the form (\ref{W_state}) always indicates the existence of TED. Uncharacterized parties in the SD can be non-participating parties in TED. The number of PP can be reduced by putting constraints on the state parameters (equality of some state parameters). 
\end{coro}
This holds because the number of non-participating parties in the entanglement scenario is equal to the number of uncharacterized parties in the steering scenario for the W-state (\ref{W_state}), i.e., ($NP = S$).

\textit{Conclusion.---} In the proposed threshold distillation protocol, fewer parties participate in achieving the objective of the quantum distillation task. Specifically, we have found that if the shared quantum state is an imperfect GHZ state, only one party is enough to distil the perfect GHZ state shared between an arbitrary number of parties and an arbitrary finite-dimensional subsystem. This feature implies that steerable correlations of such states are distillable in all steering scenarios starting from one-sided device independent ($1SDI$) to $P-1$ sided device independent ($(P-1)SDI$) where $P$ is the number of parties in the network (threshold distillation is possible upto $(P-2)SDI$). On the other hand, if the shared state in the quantum network is an imperfect W-state, then at least $P-1$ parties are required to distil the perfect W-state shared between an arbitrary number of parties. Using our protocol, this feature implies that such steerable correlations are distillable only in a $1SDI$ scenario. The proposed Krauss operators for states and assemblages can be realized using tunable beamsplitters and other linear optical elements \cite{Nery20}. We provided extensive toy examples in the supplementary materials to support our findings. The threshold concept is worth exploring for a broader range of states, communication scenarios and cryptographic settings.

\begin{acknowledgments}
\textit{Acknowledgements.---} SG acknowledges Okinawa Institute of Science and Technology, Japan, for the postdoctoral support. WJM was supported in part by the JSPS KAKENHI Grant No. 21H04880.
\end{acknowledgments}

\bibliography{Gdist
}

\end{document}




\title{Supplementary Material: Threshold ($Q, P$) Quantum Distillation}


\author{Shashank Gupta}
\email{shashankg687@gmail.com}
\affiliation{Okinawa Institute of Science and Technology Graduate University, Okinawa, Japan}

\author{William John Munro}
\affiliation{Okinawa Institute of Science and Technology Graduate University, Okinawa, Japan}

\author{Carlos Cid}
\affiliation{Okinawa Institute of Science and Technology Graduate University, Okinawa, Japan}

\maketitle 

\onecolumngrid

In this supplementary, we present a number of technical results and toy examples: In Sec. (\ref{app:GHZ_TED}), we elaborate the threshold entanglement distillation (TED) protocol for the GHZ state and give the proof of the analytical expression of state fidelity after distillation. We end the section by providing a toy example by considering a tripartite qutrit GHZ state. In Sec. (\ref{app:GHZ_TSD}), we describe, the the threshold steering distillation (TSD) protocol for the GHZ state followed by a toy example to precisely discuss the action of the TSD protocol in one-sided and two-sided device independent steering scenario. Next, we discuss the action of the TED protocol on the W-state in Sec. (\ref{app:W_TED}) and TSD protocol in Sec. (\ref{app:W_TSD}) followed by a couple of toy examples. 

\section{Threshold entanglement distillation protocol ($N, d, P, Q$) for the GHZ state. }\label{app:GHZ_TED}


Consider a quantum network consisting of $N$ copies of $P-$partite, $d-$dimensional non-maximally entangled states of the following form,

\begin{equation}
  \ket{\psi^{ini}_{GHZ}} = \sum_{i=0}^{d-1} \alpha_i \ket{\otimes_{j=0}^{P-1} i_j}_d, 
  \label{GHZ_state} \quad s.t. \sum_i|\alpha_i|^2 = 1.
\end{equation}

For such states, TED is possible using $Q=1$ to $Q=P-1$ participating parties (PP). Let us consider the case when $Q=1$

\textit{Scenario-1: (N, d, P, Q=1):} Symmetry of the GHZ-state (\ref{GHZ_state}) with respect to the parties indicate that any party can be a PP. However, we define a convention in this paper that participation starts at the end. This means that when $Q=1$ is used, the last party in the network only performs local filtering operations. In case $Q=2$, the last two parties perform such operations. Now, for the given $u^{\text{th}}$ copy, the Krauss operators of the PP are as follows:

\begin{equation}
    K_0^{P-1} = \sum_{i=0}^{d-1} \frac{\alpha_0}{\alpha_i} \ket{i}\bra{i}, \quad
    K_1^{P-1} = \sum_{i=1}^{d-1} \sqrt{1-\big(\frac{\alpha_0}{\alpha_i}\big)^2} \ket{i}\bra{i}
    \label{Krauss_GHZ_ent_1}
\end{equation}
The post-measured state of the $u^{th}$ copy after the application of the local filtering operation by the last party is
\begin{equation}
    \rho_u^{\text{dist}} = \frac{\otimes_{i=0}^{P-2} \mathcal{I}^{i} \otimes K_{o_i^u}^{P-1}. \ket{\psi^{ini}_{GHZ}}\bra{\psi^{ini}_{GHZ}}. \otimes_{i=0}^{P-2} \mathcal{I}^{i} \otimes K_{o_i^u}^{{P-1}^{\dagger}}}{\text{Tr}\big[\otimes_{i=0}^{P-2} \mathcal{I}^{i} \otimes K_{o_i^u}^{P-1}. \ket{\psi^{ini}_{GHZ}}\bra{\psi^{ini}_{GHZ}}. \otimes_{i=0}^{P-2} \mathcal{I}^{i} \otimes K_{o_i^u}^{{P-1}^{\dagger}}\big]}
\end{equation}
The success probability of the distillation protocol for the $u^{th}$ copy ($P_s^u$) is $d \alpha_0^2$. The action of the local filtering operation is independent on each copy, so the failure probability (last party gets outcome $1$ for the $u^{th}$ copy where $u \in \{1, 2, \cdots, N-1\}$) is $(1-d \alpha_0^2)^{N-1}$. This implies that the overall success probability of the TED protocol ($P_s^{\text{GHZ}}$) is $1-(1-d \alpha_0^2)^{N-1}$. Now, the final state ($\rho^{\text{dist}}_{GHZ}$) after the distillation protocol can be written as a convex combination of the perfect state ($\alpha_i = \frac{1}{\sqrt{d}}, \forall i \in \{0,d-1\})$ with probability $P_s^{\text{GHZ}}$ and the initial state ( $\ket{\psi^{ini}_{GHZ}}$) with probability $1-P_s^{\text{GHZ}}$ as follows: 

\begin{equation}
    \rho^{dist}_{GHZ} = P_s^{\text{GHZ}} \ket{\psi_{GHZ}^{\alpha_i = \frac{1}{\sqrt{d}}}}\bra{\psi_{GHZ}^{\alpha_i = \frac{1}{\sqrt{d}}}}  + (1-P_s^{\text{GHZ}}) \ket{\psi^{ini}_{GHZ}} \bra{\psi^{ini}_{GHZ}} 
\end{equation}
This state is used to determine the state fidelity as follows:

\begin{equation}
    \mathcal{F}_s (\rho^{\text{dist}}, \rho_{GHZ}^{\alpha_i = \frac{1}{\sqrt{d}}})= \Bigg[1-\frac{1}{d}(1-P_s^u)^{N-1}\Big(d-\Big(\sum_{i=0}^{d-1}\alpha_i\Big)^2\Big)\Bigg]
    \label{state_fidelity_GHZ}
\end{equation}
\textit{Scenario-2: (N, d, P, Q=2):}
As per our convention, in this case, the last two parties act as PP. There are several possibilities of the Krauss operators depending upon the permutations of the state parameters as follows:
\begin{eqnarray}
    K_0^{P-1} &=& \sum_{i=0}^{\frac{d-1}{2}} \frac{\alpha_0}{\alpha_i} \ket{i}\bra{i}, \quad
    K_1^{P-1} = \sum_{i=1}^{\frac{d-1}{2}} \sqrt{1-\big(\frac{\alpha_0}{\alpha_i}\big)^2} \ket{i}\bra{i}; \nonumber \\
     K_0^{P-2} &=& \sum_{i=\frac{d-1}{2}}^{d-1} \frac{\alpha_0}{\alpha_i} \ket{i}\bra{i}, \quad
    K_1^{P-2} = \sum_{i=\frac{d-1}{2}}^{d-1} \sqrt{1-\big(\frac{\alpha_0}{\alpha_i}\big)^2} \ket{i}\bra{i}. 
\end{eqnarray}
Several other asymmetric combinations of the Krauss operators are possible by taking unequal terms in the sum. This feature is elaborated in the toy example. Likewise, the Krauss operators for the other scenarios ($Q=3, 4, \cdots, P-1$) can be defined. The interesting finding is that the success probability in any scenario with any combination of the Krauss operators is the same ($d \alpha_0^2$), and the post-measured state is also the same. This indicates that the state fidelity value is also the same (\ref{state_fidelity_GHZ}). Hence, we claim that the performance of the above TED protocol is independent of the number of parties or the PP in the quantum network. This could be highly advantageous in experimental settings and deployed networks. 

\subsection{Toy-example (N, d = 3, P = 3)}
For the toy example, we consider N copies of tripartite, 3-dimensional non-maximally entangled states as follows,

\begin{equation}
  \ket{\psi^{ini}_{GHZ_3}} = \alpha_0 \ket{000}+\alpha_1 \ket{111}+\alpha_2 \ket{222}.
  \label{GHZ3}
\end{equation}

\textit{Scenario-1: (N, d=3, P=3, Q=1):} The Krauss operator of PP are as follows

\begin{equation}
    K_0^{2} = \ket{0}\bra{0}+\frac{\alpha_0}{\alpha_1} \ket{1}\bra{1}+\frac{\alpha_0}{\alpha_2} \ket{2}\bra{2}, \quad
    K_1^{2} = \sqrt{1-\big(\frac{\alpha_0}{\alpha_1}\big)^2} \ket{1}\bra{1}+ \sqrt{1-\big(\frac{\alpha_0}{\alpha_2}\big)^2} \ket{2}\bra{2}
    \label{Krauss_GHZ_3}
\end{equation}
The post-measured state of the $u^{th}$ copy after the application of the local filtering operation by the last party is
\begin{equation}
    \rho_u^{\text{dist}} = \frac{\mathcal{I}^{0}\otimes \mathcal{I}^{1}\otimes K_{o_i^u}^{2}. \ket{\psi^{ini}_{GHZ}}\bra{\psi^{ini}_{GHZ}}.\mathcal{I}^{0}\otimes \mathcal{I}^{1}\otimes K_{o_i^u}^{{2}^{\dagger}}}{\text{Tr}\big[\mathcal{I}^{0}\otimes \mathcal{I}^{1}\otimes K_{o_i^u}^{2}. \ket{\psi^{ini}_{GHZ}}\bra{\psi^{ini}_{GHZ}}.\mathcal{I}^{0}\otimes \mathcal{I}^{1}\otimes K_{o_i^u}^{{2}^{\dagger}}\big]}
\end{equation}
The success probability of the distillation protocol for the $u^{th}$ copy ($P_s^u$) is $3 \alpha_0^2$. The action of the local filtering operation is independent on each copy, so the failure probability (last party gets outcome $1$ for the $u^{th}$ copy where $u \in \{1, 2, \cdots, N-1\}$) is $(1-3 \alpha_0^2)^{N-1}$. This implies that the overall success probability of the TED protocol ($P_s^{\text{GHZ}}$) is $1-(1-3 \alpha_0^2)^{N-1}$. Now, the final state ($\rho^{\text{dist}}$) after the distillation protocol is as follows: 

\begin{equation}
    \rho^{dist}_{GHZ} = P_s^{\text{GHZ}} \ket{\psi_{GHZ}^{\alpha_i = \frac{1}{\sqrt{3}}}}\bra{\psi_{GHZ}^{\alpha_i = \frac{1}{\sqrt{3}}}}  + (1-P_s^{\text{GHZ}}) \ket{\psi^{ini}_{GHZ}} \bra{\psi^{ini}_{GHZ}} 
\end{equation}
The state fidelity is:

\begin{equation}
    \mathcal{F}_s (\rho^{\text{dist}}, \rho_{GHZ}^{\alpha_i = \frac{1}{\sqrt{3}}})= \Bigg[1-\frac{1}{3}(1-3 \alpha_0^2)^{N-1}\Big(3-\Big(\alpha_0+\alpha_1+\alpha_2\Big)^2 \Big)\Bigg]
    \label{state_fidelity_GHZ_3}
\end{equation}
\textit{Scenario-2: (N, d=3, P=3, Q=2):}
The Krauss operators are as follows:
\begin{equation}
    K_0^{2} =  \ket{0}\bra{0}+\frac{\alpha_0}{\alpha_1} \ket{1}\bra{1}, \quad
    K_1^{2} = \sqrt{1-\big(\frac{\alpha_0}{\alpha_1}\big)^2} \ket{1}\bra{1}; \quad
     K_0^{1} = \frac{\alpha_0}{\alpha_2} \ket{2}\bra{2}, \quad
    K_1^{1} = \sqrt{1-\big(\frac{\alpha_0}{\alpha_2}\big)^2} \ket{2}\bra{2}. 
    \label{Krauss_2_GHZ_3}
\end{equation}
Note that $K_0^2.K_0^{2^{\dagger}}+K_1^2.K_1^{2^{\dagger}} = \mathcal{I}$ and $K_0^1.K_0^{1^{\dagger}}+K_1^1.K_1^{1^{\dagger}} = \mathcal{I}$. The success probability using the above Krauss operators is also the same ($3 \alpha_0^2$) with the same post-measured state yielding the same state fidelity (\ref{state_fidelity_GHZ_3}). The fidelity asymptotically reaches one, as shown in Fig. (\ref{fidelity_GHZ_fig}).

\setcounter{figure}{0}
\renewcommand{\figurename}{FIG.}
\renewcommand{\thefigure}{S\arabic{figure}}

\begin{figure}[h!]
	\includegraphics[width=0.5\textwidth]{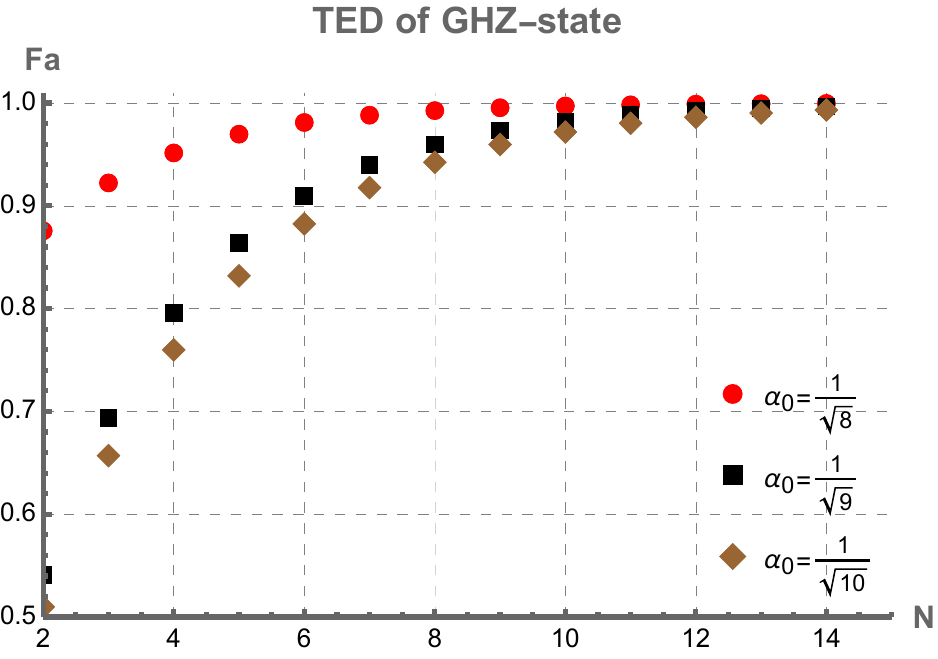}
\caption{
The vertical axis represents the state/assemblage fidelity (\ref{state_fidelity_GHZ_3}) of the GHZ state (\ref{GHZ3}), and the horizontal axis denotes the number of copies $N$ of the initial state. The plot denotes the variation of the GHZ state fidelity after TED protocol as a function of the number of states for a fixed value of the first state parameter ($\alpha_0 = \frac{1}{\sqrt{8}}, \frac{1}{\sqrt{9}}, \frac{1}{\sqrt{10}}$ for the red circle, black square and brown diamond respectively). 
}
\label{fidelity_GHZ_fig}
\end{figure}

\section{Threshold steering distillation protocol (N, d, S, P, Q) for the GHZ state}\label{app:GHZ_TSD}

Consider a quantum network consisting of $P$ parties, out of which some ($S < P$) of them are uncharacterized (untrusted), and the remaining are characterized ($P-S$). Quantum steering can arise in such networks. We start with the $N$ copies of the partially steerable assemblages derived from the GHZ state (\ref{GHZ_state}) by the action of measurements in mutually unbiased bases (MUBs) by the uncharacterized parties as follows. 

\begin{equation}
  \sigma_{a_i|x_i}^{ini} = \text{Tr}_{unch} \Big[(\otimes_{i=0}^S \mathcal{M}_{a_i|x_i} \otimes_{j=0}^{P-S} \mathcal{I}).\rho_{GHZ}^{ini}\Big]
\end{equation}
Here, we have used a simplified notation ($a_i = \{a_i\}$ and $x_i = \{x_i\}$), $a_i$ and $x_i$ denotes the string of the output and inputs if the $S$ uncharacterized parties. Depending upon the number of uncharacterized parties ($S$), a steering scenario is $1SDI$ to 
$(P-1)SDI$. For such assemblages, TSD is possible using $Q=1$ to $Q=S-1$ participating parties (PP). $Q=S$ is also a valid steering distillation scenario, but it is not a threshold distillation scenario. For a given steering scenario, we follow a convention that the starting $S$ parties are uncharacterized and out of the $P-S$ characterized parties, PP starts from the end. Let us consider the case when $Q=1$

\textit{Scenario-1: (N, d, S, P, Q=1):} As per our convention, for $Q=1$, the last party in the network only performs local filtering operations. In case $Q=2$, the last two parties perform such operations. Now, for the given $u^{\text{th}}$ copy, the Krauss operators of the PP are the same as in the case of the entanglement (\ref{Krauss_GHZ_ent_1}). Note that the major difference between steering distillation and entanglement distillation is that in the steering scenario, there are uncharacterized parties that cannot perform any filtering operations. As in TED, there are always some NPs who can be uncharacterized parties in TSD. Therefore, TSD always implies TED. 
The post-measured assemblage of the $u^{th}$ copy after the application of the local filtering operation by the last party is
\begin{equation}
    \sigma^{\text{dist}}_{a_i|x_i} = \frac{ \otimes_j \mathcal{I}_j^{NP} \otimes K_{o_i^u}^{P-1}.\sigma^{\text{ini}}_{a_i|x_i}.\otimes_j \mathcal{I}_j^{NP} \otimes K_{o_i^u}^{{P-1}^{\dagger}}}{\text{Tr}[\otimes_j \mathcal{I}_j^{NP} \otimes K_{o_i^u}^{P-1}.\rho_{ch}^{\text{ini}}. \otimes_j \mathcal{I}_j^{NP} \otimes K_{o_i^u}^{{P-1}^{\dagger}}]}
\end{equation}
Here, $\rho_{ch}^{\text{ini}}$ is the normalised conditional state on the characterized parties subsystems, i.e. $\rho_{ch}^{\text{ini}} = \text{Tr}_{unch}\rho^{ini}_{GHZ}$. The success probability of the distillation protocol for the $u^{th}$ copy ($P_s^u$) is $\text{Tr}[\otimes_j \mathcal{I}_j^{NP} \otimes K_{o_i^u}^{P-1}.\rho_{ch}^{\text{ini}}. \otimes_j \mathcal{I}_j^{NP} \otimes K_{o_i^u}^{{P-1}^{\dagger}}] = d \alpha_0^2$. The action of the local filtering operation is independent on each copy, so the failure probability (last party gets outcome $1$ for the $u^{th}$ copy of the assemblage where $u \in \{1, 2, \cdots, N-1\}$) is $(1-d \alpha_0^2)^{N-1}$. This implies that the overall success probability of the TSD protocol ($P_s^{\text{GHZ}}$) is $1-(1-d \alpha_0^2)^{N-1}$. Now, the final assemblage ($\rho^{\text{dist}}$) after the distillation protocol can be written as a convex combination of the perfect assemblage ($\alpha_i = \frac{1}{\sqrt{d}}, \forall i \in \{0,d-1\})$ with probability $P_s$ and the initial assemblage ( $\sigma_{a_i|x_i}^{ini}$) with probability $1-P_s$ as follows: 

\begin{equation}
    \sigma^{dist^{GHZ}}_{a_i|x_i} = P_s^{\text{GHZ}} \sigma_{a_i|x_i}^{GHZ}  + (1-P_s^{\text{GHZ}}) \sigma_{a_i|x_i}^{ini} 
\end{equation}

\begin{figure}[h!]
	\includegraphics[width=0.5\textwidth]{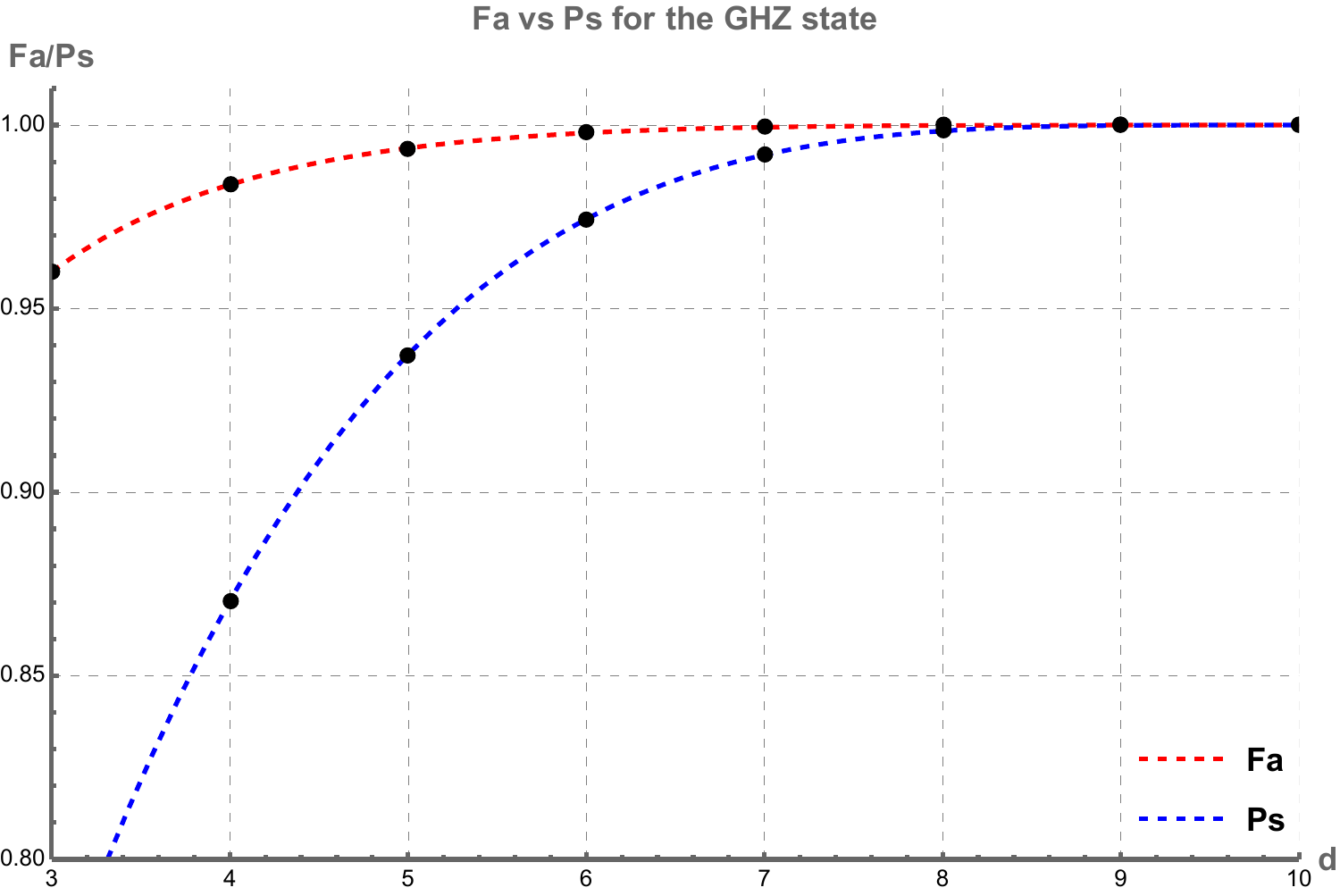}
\caption{
State/assemblage fidelity (\ref{assem_fidelity_GHZ}) (marked red colour) of the GHZ and the probability of success (blue coloured) of the distillation protocol are plotted as a function of the dimension ($d$) of the subsystem. The first state parameter ($\alpha_0$) is kept constant at ($\frac{1}{\sqrt{10}}$). The difference between the dimension and the square of the sum of the state parameters is also kept constant at 0.5.  
}
\label{Ps_fig}
\end{figure}

This assemblage is used to determine the assemblage fidelity as follows:

\begin{equation}
    \mathcal{F}_a (\{\sigma^{dist^{GHZ}}_{a_i|x_i}, \sigma_{a_i|x_i}^{GHZ}\})= \text{min}_{x_i} \sum_{a_i} \mathcal{F} (\{\sigma^{dist^{GHZ}}_{a_i|x_i}, \sigma_{a_i|x_i}^{GHZ}\}) = \Bigg[1-\frac{1}{d}(1-P_s^u)^{N-1}\Big(d-\Big(\sum_{i=0}^{d-1}\alpha_i\Big)^2 \Big)\Bigg]
    \label{assem_fidelity_GHZ}
\end{equation}
Here, minimization over the inputs of all the uncharacterized parties is performed. This expression is obtained using the numerical extension of the toy examples. Similarly, for other valid scenarios, Krauss operators for TSD will follow from the Krauss operators of TED, and the assemblage fidelity can be determined. Note that both fidelity and probability of success increase with dimension for a fixed value of the first state parameter ($\alpha_0$) and a fixed difference between the dimension and the square of the sum of the state parameters (see Fig. (\ref{Ps_fig})). 

\subsection{Toy-example (N, d = 3, S, P = 3)}
For the toy example, we consider N copies of the assemblages derived from the 4-partite, 5-dimensional, non-maximally entangled states (\ref{GHZ3}). Two kinds of steering can exist depending upon the value of $S$. One-sided device independent ($1SDI$) for $S=1$ and two-sided device independent ($2SDI$) for $S=2$. Let us first consider $1SDI$. 

\textit{Scenario-1: (N, d=3, S=1, P=3, Q=1):} 
We consider the first party to be an uncharacterized party. It performs measurement in two MUBs ($\frac{1}{\sqrt{3}} \sum_{l=0}^2 \omega^{a l + x l^2} \ket{l}, \quad (a \in \{0,2\}, x \in \{0,1\}), \omega = e^{2\pi \iota/3}$) in Hilbert space of dimension three \cite{Designolle2021}. The components of the assemblages are given by,
\begin{eqnarray}
    \sigma_{0|0}^{\text{GHZ}_3} &=& \alpha_0^2 \ket{0} \bra{0},\quad \hspace{0.45cm} 
    \sigma_{1|0}^{\text{GHZ}_3} = \alpha_1^2 \ket{1}\bra{1} ,\quad \hspace{0.45cm}
    \sigma_{2|0}^{\text{GHZ}_3} = \alpha_2^2\ket{2}\bra{2}, \nonumber \\
    \sigma_{0|1}^{\text{GHZ}_3} &=&  \ket{ \tilde{\alpha}_0} \bra{\tilde{\alpha}_0} ,\quad \hspace{0.45cm} 
     \sigma_{1|1}^{\text{GHZ}_3} = \ket{ \tilde{\alpha}_1} \bra{\tilde{\alpha}_1} ,\quad \hspace{0.45cm}
	 \sigma_{2|1}^{\text{GHZ}_3} = \ket{\tilde{\alpha}_2} \bra{\tilde{\alpha}_2}
	\label{GHZ3assemblage}
\end{eqnarray}
where $\ket{\tilde{\alpha_0}} = \frac{1}{\sqrt{3}}[\alpha_0\ket{0}+\alpha_1\ket{1}+\alpha_2\ket{2}]]$,  $\ket{\tilde{\alpha_1}} = \frac{1}{\sqrt{3}}[\alpha_0\ket{0}+ \alpha_1 \omega \ket{1}+ \alpha_2 \omega^2 \ket{2}]]$, and $\ket{\tilde{\alpha_2}} = \frac{1}{\sqrt{3}}[\alpha_0\ket{0}+ \alpha_1 \omega^2\ket{1}+\alpha_2 \omega^4\ket{2}]]$. The maximally steerable assemblage ($\sigma_{a_i|x_i}^{GHZ^3}$) corresponds to $\alpha_0 = \alpha_1 = \alpha_2 = \frac{1}{\sqrt{3}}$. After the action of the Krauss operator (\ref{Krauss_GHZ_3}), the final distilled assemblage components can be written as a convex combination of perfectly steerable assemblage with success probability ($P_s^{GHZ_3} = 1-(1-3\alpha_0^2)^{N-1})$) and initial assemblage with failure probability ($1-3\alpha_0^2)^{N-1}$). The components of the distilled assemblages are
\begin{eqnarray}
    \sigma_{0|0}^{\text{dist}_3} &=& P_s^{GHZ_3}\frac{\ket{0} \bra{0}}{3} + (1-P_s^{GHZ_3})\alpha_0^2 \ket{0} \bra{0} ,\quad \hspace{0.45cm} 
    \sigma_{1|0}^{\text{dist}_3} = P_s^{GHZ_3}\frac{\ket{1} \bra{1}}{3} + (1-P_s^{GHZ_3})\alpha_1^2 \ket{1} \bra{1} , \nonumber \\
    \sigma_{2|0}^{\text{dist}_3} &=& P_s^{GHZ_3}\frac{\ket{2} \bra{2}}{3} + (1-P_s^{GHZ_3})\alpha_2^2 \ket{2} \bra{2}, \quad \hspace{0.45cm}
    \sigma_{0|1}^{\text{dist}_3} =  P_s^{GHZ_3}\ket{\frac{1}{\sqrt{3}}_0} \bra{\frac{1}{\sqrt{3}}_0}+(1-P_s^{GHZ_3})\ket{ \tilde{\alpha}_0} \bra{\tilde{\alpha}_0} ,\nonumber \\ 
     \sigma_{1|1}^{\text{dist}_3} &=& P_s^{GHZ_3}\ket{\frac{1}{\sqrt{3}}_1} \bra{\frac{1}{\sqrt{3}}_1}+(1-P_s^{GHZ_3})\ket{ \tilde{\alpha}_1} \bra{\tilde{\alpha}_1} ,\quad \hspace{0.45cm}
	 \sigma_{2|1}^{\text{dist}_3} = P_s^{GHZ_3}\ket{\frac{1}{\sqrt{3}}_2} \bra{\frac{1}{\sqrt{3}}_2}+(1-P_s^{GHZ_3})\ket{ \tilde{\alpha}_2} \bra{\tilde{\alpha}_2}
	\label{distilGHZ3assemblage}
\end{eqnarray}
where $\ket{\frac{1}{\sqrt{3}}_0} = \frac{1}{3}[\ket{0}+\ket{1}+\ket{2}]]$,  $\ket{\frac{1}{\sqrt{3}}_1}  = \frac{1}{3}[\ket{0}+ \omega \ket{1}+ \omega^2 \ket{2}]]$, and $\ket{\frac{1}{\sqrt{3}}_2} = \frac{1}{3}[\ket{0}+ \omega^2\ket{1}+\omega^4\ket{2}]]$.

This assemblage is used to determine the assemblage fidelity as follows:

\begin{equation}
    \mathcal{F}_a (\{\sigma^{dist_3}_{a_i|x_i}, \sigma_{a_i|x_i}^{GHZ^3}\})= \text{min}_{x_i} \sum_{a_i} \mathcal{F} (\{\sigma^{dist_3}_{a_i|x_i}, \sigma_{a_i|x_i}^{GHZ^3}\}) = \Bigg[1-\frac{1}{3}(1-3 \alpha_0^2)^{N-1}\Big(3-\Big(\sum_{i=0}^{2}\alpha_i\Big)^2 \Big)\Bigg]
    \label{assem_fidelity_GHZ_3}
\end{equation}
Here, minimization over the two inputs of the uncharacterized party is performed. The expression of the assemblage fidelity is the same as the state fidelity obtained in the TED protocol.

\textit{Scenario-2: (N, d=3, S=1, P=3, Q=2):}
This is a valid scenario for steering distillation but not a valid scenario for TSD, as both the characterized parties are being used for the distillation. Using the Krauss operators (\ref{Krauss_2_GHZ_3}), the PP can distil the initial assemblage components (\ref{GHZ3assemblage}). The success probability is the same ($3 \alpha_0^2$) with the same distilled assemblages (\ref{distilGHZ3assemblage}), yielding the same assemblage fidelity (\ref{assem_fidelity_GHZ_3}).

\textit{Scenario-3: (N, d=3, S=2, P=3, Q=1):}
This is also a valid scenario for steering distillation but not a valid scenario for TSD as there is only one characterized party left for the distillation. Using the Krauss operators (\ref{Krauss_GHZ_3}), the PP can distil the initial assemblage components (36 components). The success probability comes out to be the same ($3 \alpha_0^2$), yielding the same assemblage fidelity (\ref{assem_fidelity_GHZ_3}).


\section{Threshold entanglement distillation protocol ($N, d, P, Q$) for the W state. }\label{app:W_TED}


Consider a quantum network consisting of $N$ copies of $P-$partite, non-maximally entangled states of the following form,

\begin{equation}
  \ket{\psi^{init}_{W}} = \sum_{i=0}^{P-1} \beta_i \ket{ \cdots0_{i-1}1_i0_{i+1}\cdots}, 
  \label{W_state}
\end{equation}
For such states, TED is possible using $Q=P-1$ participating parties (PP). The requirement on the number of PP can be reduced by putting constraints on the state parameters (for example, $\beta_0 = \beta_1 = \cdots = \beta_{P-2}$, requiring only two PP).

\textit{Scenario: (N, d, P, Q=P-1):} As per our convention, we consider the first party as NP. The Krauss operators of the PP are as follows:

\begin{eqnarray}
    K_0^{1} &=& \frac{\beta_{P-2}}{\beta_{P-1}}\ket{0}\bra{0} + \ket{1}\bra{1} ,\quad \hspace{0.45cm} 
    K_1^{1} = \sqrt{1-\Big(\frac{\beta_{P-2}}{\beta_{P-1}}\Big)^2} \ket{0}\bra{0} \nonumber \\
    K_0^{2} &=& \frac{\beta_{P-3}}{\beta_{P-1}}\ket{0}\bra{0} + \ket{1}\bra{1} ,\quad \hspace{0.45cm} 
    K_1^{2} = \sqrt{1-\Big(\frac{\beta_{P-3}}{\beta_{P-1}}\Big)^2} \ket{0}\bra{0} \nonumber \\
    &\vdots& \nonumber \\
    K_0^{P-1} &=& \frac{\beta_{0}}{\beta_{P-1}}\ket{0}\bra{0} + \ket{1}\bra{1} ,\quad \hspace{0.45cm} 
    K_1^{P-1} = \sqrt{1-\Big(\frac{\beta_{0}}{\beta_{P-1}}\Big)^2} \ket{0}\bra{0} \nonumber \\
    \label{Krauss_W_ent_1}
\end{eqnarray}
The post-measured state of the $u^{th}$ copy after the application of the local filtering operation by the PP parties is
\begin{equation}
    \rho_u^{\text{dist}} = \frac{\mathcal{I} \otimes_{i=1}^{P-1} K_{o_i^u}^{i}. \ket{\psi^{ini}_{W}}\bra{\psi^{ini}_{W}}. \mathcal{I} \otimes_{i=1}^{P-1} K_{o_i^u}^{{i}^{\dagger}}}{\text{Tr}\big[\mathcal{I} \otimes_{i=1}^{P-1} K_{o_i^u}^{i}. \ket{\psi^{ini}_{W}}\bra{\psi^{ini}_{W}}. \mathcal{I} \otimes_{i=1}^{P-1} K_{o_i^u}^{{i}^{\dagger}}\big]}
\end{equation}
The success probability of the distillation protocol for the $u^{th}$ copy ($P_s^u$) is $P\frac{\Pi_{i=0}^{P-1}\beta_i^2}{\beta_{P-1}^{2(P-1)}}$. The action of the local filtering operation is independent on each copy, so the failure probability (last party gets outcome $1$ for the $u^{th}$ copy where $u \in \{1, 2, \cdots, N-1\}$) is $(1-P_s^u)^{N-1}$. This implies that the overall success probability of the TED protocol ($P_s^{\text{W}}$) is $1-(1-P_s^u)^{N-1}$. Now, the final state ($\rho^{\text{dist}}$) after the distillation protocol can be written as a convex combination of the perfect state ($\beta_i = \frac{1}{\sqrt{P}}, \forall i \in \{0, P-1\})$ with probability $P_s^{\text{W}}$ and the initial state ( $\ket{\psi^{ini}_{W}}$) with probability $1-P_s^{\text{W}}$ as follows: 

\begin{equation}
    \rho^{dist}_{W} = P_s^{\text{W}} \ket{\psi_{W}^{\beta_i = \frac{1}{\sqrt{P}}}}\bra{\psi_{W}^{\beta_i = \frac{1}{\sqrt{P}}}}  + (1-P_s^{\text{W}}) \ket{\psi^{ini}_{W}} \bra{\psi^{ini}_{W}} 
\end{equation}
This state is used to determine the state fidelity as follows:

\begin{equation}
    \mathcal{F}_s (\rho^{\text{dist}}_{W}, \rho_{GHZ}^{\beta_i = \frac{1}{\sqrt{P}}})= \Bigg[1-\frac{1}{P}\Big(1-P_s^u\Big)^{N-1}\Big(P-\Big(\sum_{i=0}^{P-1}\beta_i\Big)^2 \Big)\Bigg]
    \label{state_fidelity_W}
\end{equation}

\subsection{Toy-example (N, d = 3, P = 3)}
For the toy example, we consider N copies of tripartite, non-maximally entangled W-states as follows,

\begin{equation}
  \ket{\psi^{ini}_{W_3}} = \beta_0 \ket{001}+\beta_1 \ket{010}+\beta_2 \ket{100}.
  \label{W3}
\end{equation}

\begin{figure}[h!]
	\includegraphics[width=0.5\textwidth]{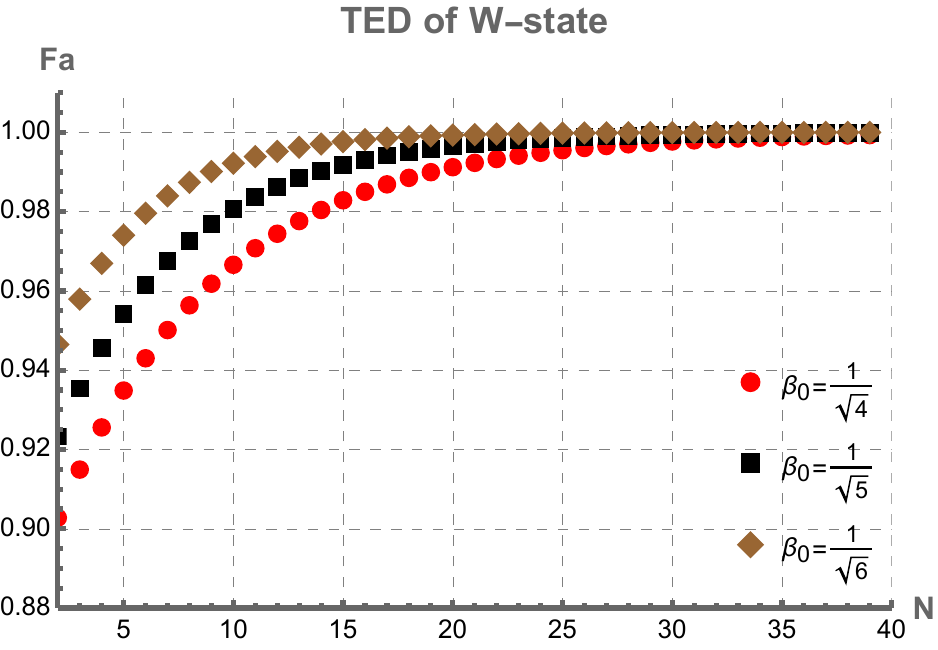}
\caption{
The vertical axis represents the state/assemblage fidelity (\ref{state_fidelity_W_3}) of the GHZ state (\ref{W3}), and the horizontal axis denotes the number of copies $N$ of the initial state. The plot denotes the variation of the W state fidelity after TED protocol as a function of the number of states for a fixed value of the first state parameter ($\beta_0 = \frac{1}{\sqrt{4}}, \frac{1}{\sqrt{5}}, \frac{1}{\sqrt{6}}$ for the red circle, black square and brown diamond respectively).
}
\label{W_fidelity_fig}
\end{figure}

\textit{Scenario: (N, d=3, P=3, Q=2):} The Krauss operator of PP are as follows

\begin{eqnarray}
    K_0^{1} &=& \frac{\beta_{1}}{\beta_{2}}\ket{0}\bra{0} + \ket{1}\bra{1} ,\quad \hspace{0.45cm} 
    K_1^{1} = \sqrt{1-\Big(\frac{\beta_{1}}{\beta_{2}}\Big)^2} \ket{0}\bra{0} \nonumber \\
    K_0^{2} &=& \frac{\beta_{0}}{\beta_{2}}\ket{0}\bra{0} + \ket{1}\bra{1} ,\quad \hspace{0.45cm} 
    K_1^{2} = \sqrt{1-\Big(\frac{\beta_{0}}{\beta_{2}}\Big)^2} \ket{0}\bra{0} 
    \label{Krauss_W_3}
\end{eqnarray}
The post-measured state of the $u^{th}$ copy after the application of the local filtering operation by the PPs is
\begin{equation}
    \rho_u^{\text{dist}} = \frac{\mathcal{I}^{0}\otimes K_{o_i^u}^{1}\otimes K_{o_i^u}^{2}. \ket{\psi^{ini}_{W_3}}\bra{\psi^{ini}_{W_3}}.\mathcal{I}^{0}\otimes K_{o_i^u}^{{1}^{\dagger}}\otimes K_{o_i^u}^{{2}^{\dagger}}}{\text{Tr}\big[\mathcal{I}^{0}\otimes K_{o_i^u}^{1}\otimes K_{o_i^u}^{2}. \ket{\psi^{ini}_{W_3}}\bra{\psi^{ini}_{W_3}}.\mathcal{I}^{0}\otimes K_{o_i^u}^{{1}^{\dagger}}\otimes K_{o_i^u}^{{2}^{\dagger}}\big]}
\end{equation}
The success probability of the distillation protocol for the $u^{th}$ copy ($P_s^u$) is $3 \frac{\beta_0^2 \beta_1^2}{\beta_2^2}$. The action of the local filtering operation is independent on each copy, so the failure probability (last party gets outcome $1$ for the $u^{th}$ copy where $u \in \{1, 2, \cdots, N-1\}$) is $(1-3 \frac{\beta_0^2 \beta_1^2}{\beta_2^2})^{N-1}$. This implies that the overall success probability of the TED protocol ($P_s^{\text{W}_3}$) is $1-(1-3 \frac{\beta_0^2 \beta_1^2}{\beta_2^2})^{N-1}$. Now, the final state ($\rho^{\text{dist}}$) after the distillation protocol is as follows: 

\begin{equation}
    \rho^{dist}_{W} = P_s^{\text{W}_3} \ket{\psi_{W}^{\beta_i = \frac{1}{\sqrt{3}}}}\bra{\psi_{W}^{\beta_i = \frac{1}{\sqrt{3}}}}  + (1-P_s^{\text{W}_3}) \ket{\psi^{ini}_{W_3}} \bra{\psi^{ini}_{W_3}} 
\end{equation}
The state fidelity is:

\begin{equation}
    \mathcal{F}_s (\rho^{\text{dist}}, \rho_{W}^{\beta_i = \frac{1}{\sqrt{3}}})= \Bigg[1-\frac{1}{3}(1-3 \frac{\beta_0^2 \beta_1^2}{\beta_2^2})^{N-1}\Big(\Big(\beta_0+\beta_1+\beta_2\Big)^2 - 3\Big)\Bigg]
    \label{state_fidelity_W_3}
\end{equation}
The fidelity asymptotically reaches one, as shown in Fig. (\ref{W_fidelity_fig}).

\section{Steering distillation protocol (N, d, S, P, Q) for the W state}\label{app:W_TSD}


 We start with the $N$ copies of the partially steerable assemblages derived from the non-maximally entangled W state (\ref{W_state}) by the action of measurements in mutually unbiased bases (MUBs) by the uncharacterized parties as follows \cite{Designolle2021}. 

\begin{equation}
  \sigma_{a_i|x_i}^{ini} = \text{Tr}_{unch} \Big[(\otimes_{i=0}^S \mathcal{M}_{a_i|x_i} \otimes_{j=0}^{P-S} \mathcal{I}).\rho_{W}^{ini}\Big]
\end{equation}
Here, we have used a simplified notation ($a_i = \{a_i\}$ and $x_i = \{x_i\}$), $a_i$ and $x_i$ denotes the string of the output and inputs of the $S$ uncharacterized parties. We noticed that for W-state (\ref{W_state}), the TED protocol demands $P-1$ PPs. Therefore, for such states, TSD is not possible. However, TSD might be possible if there is a constraint on the state parameters. Here, we consider the steering distillation for the one-sided device-independent scenario as follows:

\textit{Scenario-1: (N, d, S, P, Q=P-1):} As per our convention, we consider the first party as an uncharacterized party and the second party onwards as a characterized party. Now, for the given $u^{\text{th}}$ copy, the Krauss operators of the PP are the same as in the case of the entanglement (\ref{Krauss_W_ent_1}). Note that the major difference between steering distillation and entanglement distillation is that in the steering scenario, there are uncharacterized parties that cannot perform any filtering operations. As in TED, there are always some NPs who can be uncharacterized parties in TSD. Therefore, TSD always implies TED. 

The post-measured assemblage of the $u^{th}$ copy after the application of the local filtering operation by the PP is
\begin{equation}
    \sigma^{\text{dist}}_{a_i|x_i} = \frac{ \mathcal{I}_0^{NP} \otimes_{i=1}^{P-1} K_{o_i^u}^{i}.\sigma^{\text{ini}}_{a_i|x_i}. \mathcal{I}_0^{NP} \otimes_{i=1}^{P-1} K_{o_i^u}^{{i}^{\dagger}}}{\text{Tr}[ \mathcal{I}_0^{NP} \otimes_{i=0}^{P-1} K_{o_i^u}^{i}.\rho_{ch}^{\text{ini}}. \mathcal{I}_0^{NP} \otimes_{i=1}^{P-1} K_{o_i^u}^{{i}^{\dagger}}]}
\end{equation}
Here, $\rho_{ch}^{\text{ini}}$ is the normalised conditional state on the characterized parties subsystems, i.e. $\rho_{ch}^{\text{ini}} = \text{Tr}_{unch}\rho^{ini}_{W}$. The success probability of the distillation protocol for the $u^{th}$ copy ($P_s^u$) is $\text{Tr}[ \mathcal{I}_0^{NP} \otimes_{i=1}^{P-1} K_{o_i^u}^{i}.\rho_{ch}^{\text{ini}}. \mathcal{I}_0^{NP} \otimes_{i=1}^{P-1} K_{o_i^u}^{{i}^{\dagger}}] = P \frac{\Pi_{i=0}^{P-1}\beta_i^2}{\beta_{P-1}^{2(P-1)}}$. The action of the local filtering operation is independent on each copy, so the failure probability (last party gets outcome $1$ for the $u^{th}$ copy of the assemblage where $u \in \{1, 2, \cdots, N-1\}$) is $(1-P_s^u)^{N-1}$. This implies that the overall success probability of the TSD protocol ($P_s^{\text{W}}$) is $1-(1-P_s^u)^{N-1}$. Now, the final assemblage ($\sigma^{\text{dist}}$) after the distillation protocol can be written as a convex combination of the perfect assemblage ($\beta_i = \frac{1}{\sqrt{P}}, \forall i \in \{0, P-1\})$ with probability $P_s^{\text{W}}$ and the initial assemblage ( $\sigma_{a_i|x_i}^{ini}$) with probability $1-P_s^{\text{W}}$ as follows: 

\begin{equation}
    \sigma^{dist^{W}}_{a_i|x_i} = P_s^{\text{W}} \sigma_{a_i|x_i}^{W}  + (1-P_s^{\text{W}}) \sigma_{a_i|x_i}^{ini} 
\end{equation}
This assemblage is used to determine the assemblage fidelity as follows:

\begin{equation}
    \mathcal{F}_a (\{\sigma^{dist^{W}}_{a_i|x_i}, \sigma_{a_i|x_i}^{W}\})= \text{min}_{x_i} \sum_{a_i} \mathcal{F} (\{\sigma^{dist^{W}}_{a_i|x_i}, \sigma_{a_i|x_i}^{W}\}) = \Bigg[1-\frac{1}{P}(1- P_s^u)^{N-1}\Big(P-\Big(\sum_{i=0}^{P-1}\beta_i\Big)^2 \Big)\Bigg]
    \label{assem_fidelity_W}
\end{equation}
Here, minimization over the inputs of all the uncharacterized parties is performed. This expression is obtained using the numerical extension of the toy examples.


\subsection{Toy-example (N, d = 2, S=1, P = 3)}
For the toy example, we consider N copies of the assemblages derived from the tripartite, non-maximally entangled states (\ref{W3}). Two kinds of steering can exist depending upon the value of $S$. One-sided device independent ($1SDI$) for $S=1$ and two-sided device independent ($2SDI$) for $S=2$. Since $P-1$ PPs are required for W-state distillation, we only consider the $1SDI$ scenario. 

\textit{Scenario-1: (N, d=3, S=1, P=3, Q=2):} 
We consider the first party to be an uncharacterized party. It performs measurement in two MUBs ($\frac{1}{\sqrt{2}} \sum_{l=0}^1 \omega^{a l + x l^2} \ket{l}, \quad (a \in \{0,1\}, x \in \{0,1\}), \omega = e^{2\pi \iota/2}$) in Hilbert space of dimension two \cite{Designolle2021}. The two MUBs are computational and on a Hadamard basis. The components of the assemblages are given by,
\begin{eqnarray}
    \sigma_{0|0}^{\text{W}_3} &=& \beta_0^2 + \beta_1^2 \ket{w_0} \bra{w_0},\quad \hspace{0.45cm} 
    \sigma_{1|0}^{\text{W}_3} = \beta_2^2 \ket{00}\bra{00} ,\nonumber \\
    \sigma_{0|1}^{\text{W}_3} &=& \frac{1}{2}\ket{w_1^+}\bra{w_1^+}, \quad \hspace{0.45cm} 
    \sigma_{1|1}^{\text{W}_3} =  \frac{1}{2}\ket{w_1^-} \bra{w_1^-} 
	\label{W3assemblage}
\end{eqnarray}
where $\ket{w_0} = \frac{1}{\sqrt{\beta_0^2+\beta_1^2}}[\beta_0\ket{01}+\beta_1\ket{10}]$,  $\ket{w_1^{\pm}} = [\beta_2\ket{00}\pm \beta_0 \ket{01}\pm \beta_1 \ket{10}]$. The maximally steerable assemblage ($\sigma_{a_i|x_i}^{W^3}$) corresponds to $\beta_0 = \beta_1 = \beta_2 = \frac{1}{\sqrt{3}}$. After the action of the Krauss operator (\ref{Krauss_W_3}), the final distilled assemblage components can be written as a convex combination of the perfectly steerable assemblage with success probability ($P_s^{W_3} = 1-(1-3\frac{\beta_0^2 \beta_1^2}{\beta_2^2})^{N-1})$) and initial assemblage with failure probability ($1-3\frac{\beta_0^2 \beta_1^2}{\beta_2^2})^{N-1}$). The components of the distilled assemblages are
\begin{eqnarray}
    \sigma_{0|0}^{\text{dist}_3} &=& P_s^{W_3}\frac{2}{3} \ket{\frac{1}{\sqrt{3}}_0}\bra{\frac{1}{\sqrt{3}}_0} + (1-P_s^{W_3})\beta_0^2 + \beta_1^2 \ket{w_0} \bra{w_0} ,\quad \hspace{0.45cm} 
    \sigma_{1|0}^{\text{dist}_3} = P_s^{W_3}\frac{1}{3}\ket{00}\bra{00} + (1-P_s^{W_3})\beta_2^2 \ket{00}\bra{00} , \nonumber \\
    \sigma_{0|1}^{\text{dist}_3} &=&  P_s^{W_3}\frac{1}{2}\ket{\frac{1}{\sqrt{3}}_1^+}\bra{\frac{1}{\sqrt{3}}_1^+}+(1-P_s^{W_3})\frac{1}{2}\ket{w_1^+}\bra{w_1^+} ,\quad 
     \sigma_{1|1}^{\text{dist}_3} = P_s^{W_3}\frac{1}{2}\ket{\frac{1}{\sqrt{3}}_1^+}\bra{\frac{1}{\sqrt{3}}_1^+}+(1-P_s^{W_3})\frac{1}{2}\ket{w_1^+}\bra{w_1^+}
	\label{distilW3assemblage}
\end{eqnarray}
where $\ket{\frac{1}{\sqrt{3}}_0} = \frac{1}{\sqrt{2}}[\ket{01}+\ket{10}]$, $\ket{\frac{1}{\sqrt{3}}_1^{\pm}} = [\frac{1}{\sqrt{3}}(\ket{00}\pm \ket{01}\pm \ket{10})]$.

This assemblage is used to determine the assemblage fidelity as follows:

\begin{equation}
    \mathcal{F}_a (\{\sigma^{dist_3}_{a_i|x_i}, \sigma_{a_i|x_i}^{W^3}\})= \text{min}_{x_i} \sum_{a_i} \mathcal{F} (\{\sigma^{dist_3}_{a_i|x_i}, \sigma_{a_i|x_i}^{W^3}\}) = \Bigg[1-\frac{1}{3}(1-3 \frac{\beta_0^2 \beta_1^2}{\beta_2^2})^{N-1}\Big(3-\Big(\sum_{i=0}^{2}\beta_i\Big)^2 \Big)\Bigg]
    \label{assem_fidelity_W_3}
\end{equation}
Here, minimization over the two inputs of the uncharacterized party is performed. The expression of the assemblage fidelity is the same as the state fidelity obtained in the TED protocol.



\bibliography{Gdist}